\documentclass[]{spie}

\usepackage{bytefield}
\usepackage{amsmath,amsfonts,amssymb}
\usepackage{graphicx}
\usepackage{subcaption} 
\usepackage[colorlinks=true, allcolors=blue]{hyperref}
\usepackage{siunitx}
\usepackage{array}
\usepackage{booktabs}
\usepackage{cite}
\usepackage{tikz}
\usetikzlibrary{arrows.meta,positioning,shapes.geometric}

\usepackage{float}
\usepackage{placeins}
\usepackage{multirow}
\usepackage{parskip}

\usepackage{xcolor}
\usepackage{afterpage}

\title{Development of an FPGA-Based Readout System and Functional Characterization of a Custom NSX1 ASIC for X-ray instrumentation}

\author[a]{Nabin Poudyal}
\author[b]{Wayne Baumgartner}
\author[a]{Sukanya Chakrabarti}
\author[c,d]{Gianluigi De Geronimo}
\author[a]{Wesley Hodge}
\author[a]{Peter Jenke}
\author[b]{Paul Ray}
\author[e]{Jarred Roberts}
\author[b]{Eric Wulf}

\affil[a]{University of Alabama Huntsville, 301 Sparkman Drive, Alabama 35899, USA}
\affil[b]{U.S. Naval Research Laboratory, 4555 Overlook Ave. S.W. Washington, DC 20375, USA}
\affil[c]{DG Circuits, dgcircuits.com, Syosset, NY 11973, USA}
\affil[d]{Electrical and Computer Engineering, Stony Brook University, Stony Brook, NY 11794, USA}
\affil[e]{University of California San Diego, 9500 Gilman Dr. La Jolla, CA 92093, USA}

\authorinfo{Further author information: (Send correspondence to N.P.)\\N.P.: E-mail: nabin.poudyal@uah.edu}

\begin{document}
\maketitle

\begin{abstract}
We developed an FPGA-based readout system for a front-end ASIC (NSX1) with 64 channels to read out a large-area silicon drift detector used in X-ray detection instruments such as the Wide Field Monitor module for the STROBE-X mission concept. The FPGA acts as a central controller for the NSX1 configuration, implementing trigger handling and event sequencing, precise event timestamping, control of an external ADC for digitization, and data transfer to the host PC. The system enables comprehensive ASIC characterization; here, we demonstrate preliminary ASIC functionality and characterization tests using a calibrated pulser for two programmable linear gain responses of 375 mV/fC and 750 mV/fC.
 These preliminary results validate the low-noise performance of the NSX1 ASIC and demonstrate successful integration of the DAQ system. 
\end{abstract}




\section{Introduction}
\label{sec:introduction}

STROBE-X~\cite{ray2019strobexxraytimingspectroscopy} is a proposed NASA probe-class observatory for X-ray spectroscopy and timing mission. It is composed of three instruments: the Low Energy Modular Array (LEMA), the High Energy Modular Array (HEMA), and the Wide Field Monitor (WFM). This paper reports preliminary FPGA firmware development and readout design for the WFM instrument. The total sensitive area of the WFM instrument is 1453~cm$^2$ and is made up of 8 separate modules. Each WFM module contains 4 linear silicon drift detectors (SDD), and each SDD is divided into two halves along its drift direction. A WFM SDD is approximately 70~mm $\times$ 64~mm in size. A half-SDD has 384 anode pads at a pitch of 169~\textmu m to collect the charge generated by X-ray interactions and is read out by 6 custom NSX1 ASICs, each with 64 channels \cite{degeronimo2024frontendasicstrobexhema}.

The NSX1 readout system consists of the ASIC itself, an FPGA-based controller, an optional external ADC, a USB FIFO link to the host PC, and Python-based data acquisition software called Python-GSE. The NSX1, a 64-channel low-noise, low-power ASIC, is optimized for the WFM instrument and has an on-chip charge amplifier, a shaper with a baseline stabilizer, an event discriminator, a peak detector, and control logic with configuration registers. The NSX1 integrates an on-chip ADC for digitization of the peak-detect output. The hardware for the FPGA controller is the Lattice LFD2NX-40-9BG256I (built with the Radiant v2024.2 toolchain) with about 39k logic cells, 192 user I/O pins, and an external SPI flash configuration \cite{lattice_certusnx}. It generates clock and control signals to configure, control, and read the NSX1. The Python-GSE software is responsible for setting FPGA control registers and ASIC configuration via the SPI protocol, as well as reading out and decoding the event packets and saving them to disk. The FPGA firmware is responsible for data packing, event building, and sending data to the host PC. A Cypress FX2LP USB 2.0 High-Speed 16-bit slave-FIFO interface provides buffered high-throughput data transfer and acts as a bridge between the host PC and the FPGA \cite{cypress_fx2lp_cy7c68013a}.

The complete readout system is divided into two parts: the NSX1 ASIC as the device under test (DUT) and the FPGA, USB-FIFO and Python-GSE together as the measurement instrument. Before characterizing the ASIC, the measurement instrument itself must be verified to behave correctly. The following characterization metrics are reported in this paper. On the FPGA side, we report RTL and gate-level functional verification, static and dynamic power, timing closure and the maximum operating clock, signal integrity and bit-error rate (BER) on the USB and ASIC interfaces, and readout dead time and rate capability. On the ASIC side, we report configuration and protocol verification, noise and equivalent noise charge (ENC) studies, baseline and pulse-output stability versus time, pulser DAC and charge-injection response, threshold DAC linearity and trim verification, pulser gain linearity, analog gain, peaking time, and pulse shape, and finally readout timing, dead time, and rate.


\section{Readout System Design Requirement}
\label{sec:requirements}

STROBE-X performs X-ray spectroscopy and timing over a broad energy range, and the WFM instrument covers 2 to 50~keV (up to 80~keV in extended mode) with a low-energy threshold of about 470~eV. The WFM uses linear silicon drift detectors (SDDs)~\cite{Bertuccio_2016}, in which photoelectron drift takes up to 2.2~$\mu$s to produce a 160~ns current pulse at the anode. The detector geometry and the NSX1 front-end ASIC are inherited from the WFM design~\cite{degeronimo2024frontendasicstrobexhema}. The parameters that are included in the firmware design are summarized in Table~\ref{tab:detector_asic_requirements}.
NSX1 is a 64-channel front-end ASIC. Each channel contains a dual-stage charge-sensitive amplifier, a shaper with baseline stabilization, a discriminator, and a peak detector. During acquisition mode, when an event is detected, it produces an event flag (FL) after a $\sim$400~ns delay, indicating the start of the ASIC data readout\cite{nsx1_datasheet}. Once the ASIC enters the readout mode, each rising edge of the ASIC clock generates an internal readout token which reads all channel data sequentially and digitizes it by an on-chip ADC or by an external ADC. The ASIC configuration parameters, such as analog gain, peaking time, threshold trim, pulser injection, neighbor logic, etc., are programmed through a 1024-bit SPI configuration bitstream.
 
On the FPGA side, the firmware drives the ASIC interface (SDI, RW, CS, CK, EN) and writes the 1024-bit configuration bitstream to each ASIC before acquisition. The firmware implements finite state machines that transition through configuration, acquisition, and readout. During readout mode, the firmware drives the ADC paths and reads out data from all the channels. It formats the data into a packet of four words which includes timestamp, channel ID, ADC data, and ASIC ID. The packets are then streamed to the host through the Cypress FX2LP USB interface. Currently, the firmware supports two ASICs in this work but is being designed to scale to the six ASICs per half-SDD used in the WFM module. On the PC side, Python-GSE writes FPGA control registers, sends ASIC SPI commands, reads back the USB packets from the FPGA, parses them, and converts them to HDF5.

Table~\ref{tab:detector_asic_requirements} summarizes the detector and ASIC-level requirements taken as given for this work, and Table~\ref{tab:fpga_daq_requirements} summarizes the FPGA firmware and DAQ requirements to implement the requirements. The full WFM and HEMA front-end specification is given in~\cite{degeronimo2024frontendasicstrobexhema}.

\begin{table}[!t]
\centering
\caption{Detector and ASIC-level inputs. Detector values are for the WFM front-end from~\cite{ray2019strobexxraytimingspectroscopy} and the NSX1 ASIC architecture values are from~\cite{degeronimo2024frontendasicstrobexhema}.}
\label{tab:detector_asic_requirements}
\small
\begin{tabular}{>{\centering\arraybackslash}p{0.02\linewidth}| >{\raggedright\arraybackslash}p{0.22\linewidth} >{\raggedright\arraybackslash}p{0.27\linewidth} >{\raggedright\arraybackslash}p{0.30\linewidth}}
\toprule
\textbf{} & \textbf{Category} & \textbf{Requirement} & \textbf{Explanation} \\
\midrule

\multirow{6}{*}{\centering\rotatebox[origin=c]{90}{\textbf{Detector Level}}}
& X-ray energy range          & 2--50~keV (up to 80~keV, extended mode) & Sets the ASIC gain and ADC code range. \\
& Energy threshold            & $\sim$470~eV                            & Sets the threshold-DAC resolution and trim range. \\
& Charge drift time           & $\leq 2.2~\mu\text{s}$                  & Minimum acquire window the firmware must keep open. \\
& Charge collection time      & $\leq 160~\text{ns}$                    & Effect on selection of a peaking time. \\

\midrule

\multirow{8}{*}{\centering\rotatebox[origin=c]{90}{\textbf{ASIC Level}}}
& Architecture      & 64-channel integrated front-end & Includes amplification, shaping, discrimination, peak detection, and ADC conversion \\
& Operation         & Configuration, Acquisition, Readout modes & Two-phase operation controlled via digital interface \\
& Programmability   & Gain, peaking time, threshold trim & Optimization of noise, dynamic range, and channel uniformity \\
& Event Processing  & Peak detection with event flag (FL) & Stores peak amplitude and signals valid events \\
& Readout           & Token-based multiplexing & Channel selection based on the clock sequence \\
& ADC / Output      & Internal or external ADC & Supports flexible digitization architectures \\
& Advanced Features & Neighbor and trigger-all modes & Enables multi-channel and inter-chip event correlation \\
& Calibration       & Internal pulser and monitoring & Supports calibration and system diagnostics \\

\bottomrule
\end{tabular}
\end{table}

\begin{table}[!t]
\centering
\caption{FPGA firmware and DAQ software requirements derived from the detector and ASIC-level inputs.}
\label{tab:fpga_daq_requirements}
\small
\begin{tabular}{>{\centering\arraybackslash}p{0.02\linewidth}| >{\raggedright\arraybackslash}p{0.22\linewidth} >{\raggedright\arraybackslash}p{0.27\linewidth} >{\raggedright\arraybackslash}p{0.30\linewidth}}
\toprule
\textbf{} & \textbf{Category} & \textbf{Requirement} & \textbf{Explanation} \\
\midrule

\multirow{8}{*}{\centering\rotatebox[origin=c]{90}{\textbf{FPGA Level}}}
& Control & Full ASIC interface management & Drives SDI, RW, CS, CK, EN signals \\
& Timing & Deterministic sequencing & Ensures correct state transitions \\
& Configuration & Serial configuration engine & Programs full ASIC register \\
& State Machines & Acquisition and readout FSMs & Controls event capture and data shift flow \\
& Readout Logic & Channel scan and FL decoding & Reconstructs channel data and event validity \\
& ADC Handling & Internal/external ADC support & Interface internal/external ADC \\
& Data Handling & Packet formation and buffering & Transfer data using FIFO structures \\
& Scalability & Multi-ASIC support & Enables expansion to larger detector systems \\

\midrule

\multirow{4}{*}{\centering\rotatebox[origin=c]{90}{\textbf{DAQ Level}}}
& Data Processing & Structured event building & Combines channel, amplitude, and metadata \\
& Throughput & Continuous high-rate acquisition & Prevents event loss under high event rates \\
& Monitoring & Real-time diagnostics & Tracks temperature and bandgap reference \\
& Data Integrity & Error detection and logging & Captures overflow and ADC-related errors \\

\bottomrule
\end{tabular}
\end{table}

\FloatBarrier


\section{FPGA Readout System Architecture}
\label{sec:fpga_architecture}

The readout system is composed of two PCBs: a first analog board called SPHINX, and a second digital board called SPICE.
The SPHINX board holds the two NSX1 ASICs and two 16-bit AD7980 ADCs for digitization. It also supplies low-noise analog power and filtering. It provides the pads to directly wire-bond the SDD detector anodes to the ASIC input pads. 
The SPICE board hosts a Lattice Certus-NX FPGA, which runs the sequencing and digital logic. It also generates digital power supplies and filtering for the FPGA. The SPICE board also hosts a Cypress EZ-USB FX2 controller, which acts as a bridge between the FPGA and the DAQ computer.

\subsection{Hardware Platform}
\label{subsec:hardware_platform}

\begin{figure}[!htbp]
  \centering
  \includegraphics[width=1\linewidth]{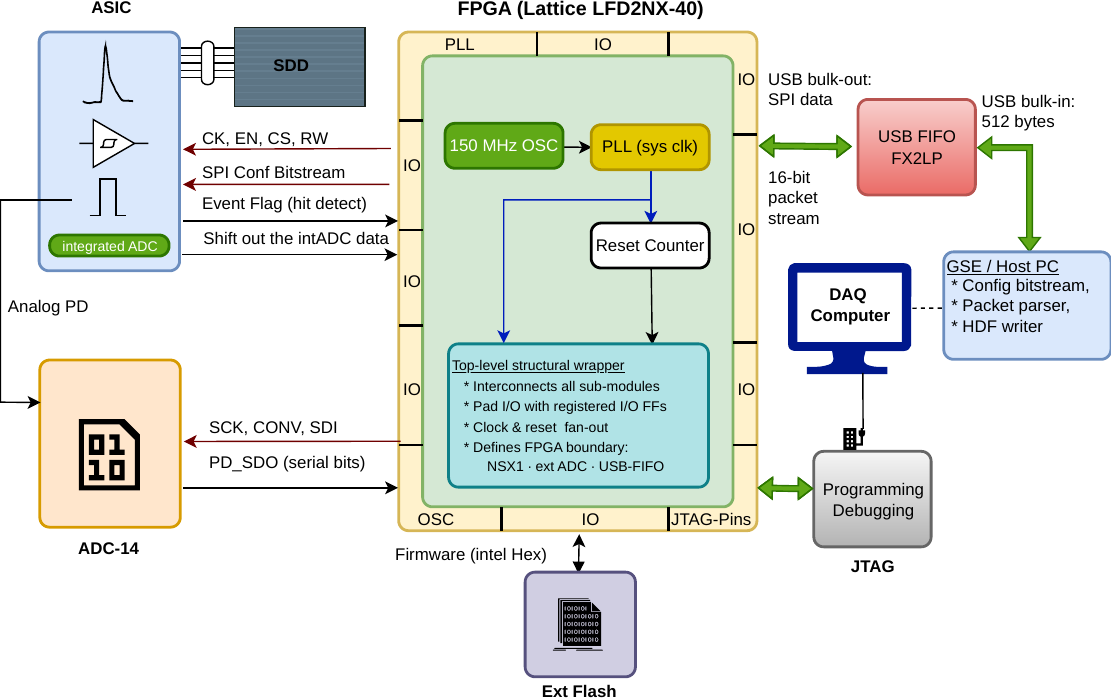}
  \caption{FPGA hardware platform and external interfaces.}
  \label{fig:fpga_hardware_platform}
\end{figure}

We selected the Lattice LFD2NX-40-9BG256I FPGA for its low power consumption. 
The FPGA has an oscillator, three PLLs for clock management, about 39k logic cells, and 192 user I/O pins.
A system clock of 50~MHz is generated from a 150~MHz internal oscillator and an on-chip PLL, and is used as the reference for all FPGA timing measurements. 
The FPGA logic, described in Verilog, was implemented using the Radiant 2024.2 toolchain (Table~\ref{tab:hardware_platform}),
and the bitstream is deployed from either an external SPI flash or directly using JTAG programming. 
The ASIC and USB interface clocks are generated using the system clock. 
Figure~\ref{fig:fpga_hardware_platform} illustrates a top-level block diagram of the FPGA readout system architecture. 
It shows the external interfaces: the SPI configuration and readout signals to each NSX1 (SDI, CS, RW, CK, EN, FL), the external ADC signals (CONV, SCLK, SDO), the pulser strobe, the JTAG/programming pins, and the 16-bit slave-FIFO bus to the FX2LP. 
Table~\ref{tab:fpga_asic_boundary} lists the signals between the FPGA and the ASIC. All timing and signal-integrity measurements in this paper are taken at these pins.

\begin{table}[!htbp]
  \centering
  \caption{Physical signals defining the FPGA-to-NSX1 boundary.}
  \label{tab:fpga_asic_boundary}
  \small
  \begin{tabular}{|p{0.18\linewidth}|p{0.28\linewidth}|p{0.42\linewidth}|}
    \hline
    \textbf{Direction} & \textbf{Signal (per channel)} & \textbf{Purpose} \\
    \hline
    \multicolumn{3}{|l|}{\textit{FPGA $\leftrightarrow$ NSX1 ASIC ($\times 2$ instances)}} \\
    \hline
    FPGA $\rightarrow$ ASIC & \texttt{nrlasic\{0,1\}\_sdi}      & SPI configuration serial data \\
    FPGA $\rightarrow$ ASIC & \texttt{nrlasic\{0,1\}\_cs}       & Chip select \\ 
    FPGA $\rightarrow$ ASIC & \texttt{nrlasic\{0,1\}\_rw}       & SPI read/write \\ 
    FPGA $\rightarrow$ ASIC & \texttt{nrlasic\{0,1\}\_ck}       & Readout clock (LVDS) \\ 
    FPGA $\rightarrow$ ASIC & \texttt{nrlasic\{0,1\}\_en}       & Readout enable (LVDS) \\ 
    ASIC $\rightarrow$ FPGA & \texttt{nrlasic\{0,1\}\_flag}     & Event flag FL (LVDS) \\ \hline
    \multicolumn{3}{|l|}{\textit{FPGA $\leftrightarrow$ external ADC ($\times 2$ instances)}} \\
    \hline
    FPGA $\rightarrow$ ADC  & \texttt{adc\{0,1\}\_conv}         & Start-convert strobe \\ 
    FPGA $\rightarrow$ ADC  & \texttt{adc\{0,1\}\_sclk}         & ADC serial clock \\ 
    ADC  $\rightarrow$ FPGA & \texttt{adc\{0,1\}\_sdo}          & Digitized peak amplitude \\ \hline
  \end{tabular}
\end{table}

\begin{table}[!htbp]
  \centering
  \caption{Hardware platform summary.}
  \label{tab:hardware_platform}
  \small
  \begin{tabular}{|p{0.20\linewidth}|p{0.30\linewidth}|p{0.38\linewidth}|}
    \hline
    \textbf{Item}         & \textbf{Value}                              & \textbf{Role in characterization} \\ \hline
    FPGA device           & Lattice LFD2NX-40-9BG256I (Radiant v2024.2) &  Hosts all firmware logic, toolchain, and synthesis results \\
    FPGA board revision   & SPICEv2 Rev.\ 1                              &  Maps firmware signals to physical board pins  \\
    NSX1 board revision   & SPHINX1 board Rev.\ 1                        &  Sets the analog paths between detector, pulser, and ASIC \\
    System clock frequency& 50~MHz (internal oscillator, pll)           &  Reference clock for every timing measurement \\
    USB FIFO interface    & Cypress FX2LP, USB 2.0 HS, 16-bit bus       &  High throughput, buffering and USB transfer \\
    Integrated ADC mode   & NSX1 on-chip ADC, serial PD/PDN output      &  Integrate digitization path \\
    External ADC path     & AD7980 (AD7980SRMZ-EP-RL7)                  &  Alternative: digitization path \\
    Pulser source         & NSX1 internal pulser DAC                    &  Calibrates charge injection \\ 
    Host computer and OS  & Ubuntu 22.04, x86-64, Python 3.11.5, \texttt{asic\_gse} vX.Y &    \\
    \hline
  \end{tabular}
\end{table}

\FloatBarrier


\section{HDL Design and Firmware Architecture}
\label{sec:hdl_methodology}

\begin{figure}[!htbp]
  \centering
  \begin{subfigure}{\linewidth}
    \centering
    \includegraphics[width=0.9\linewidth]{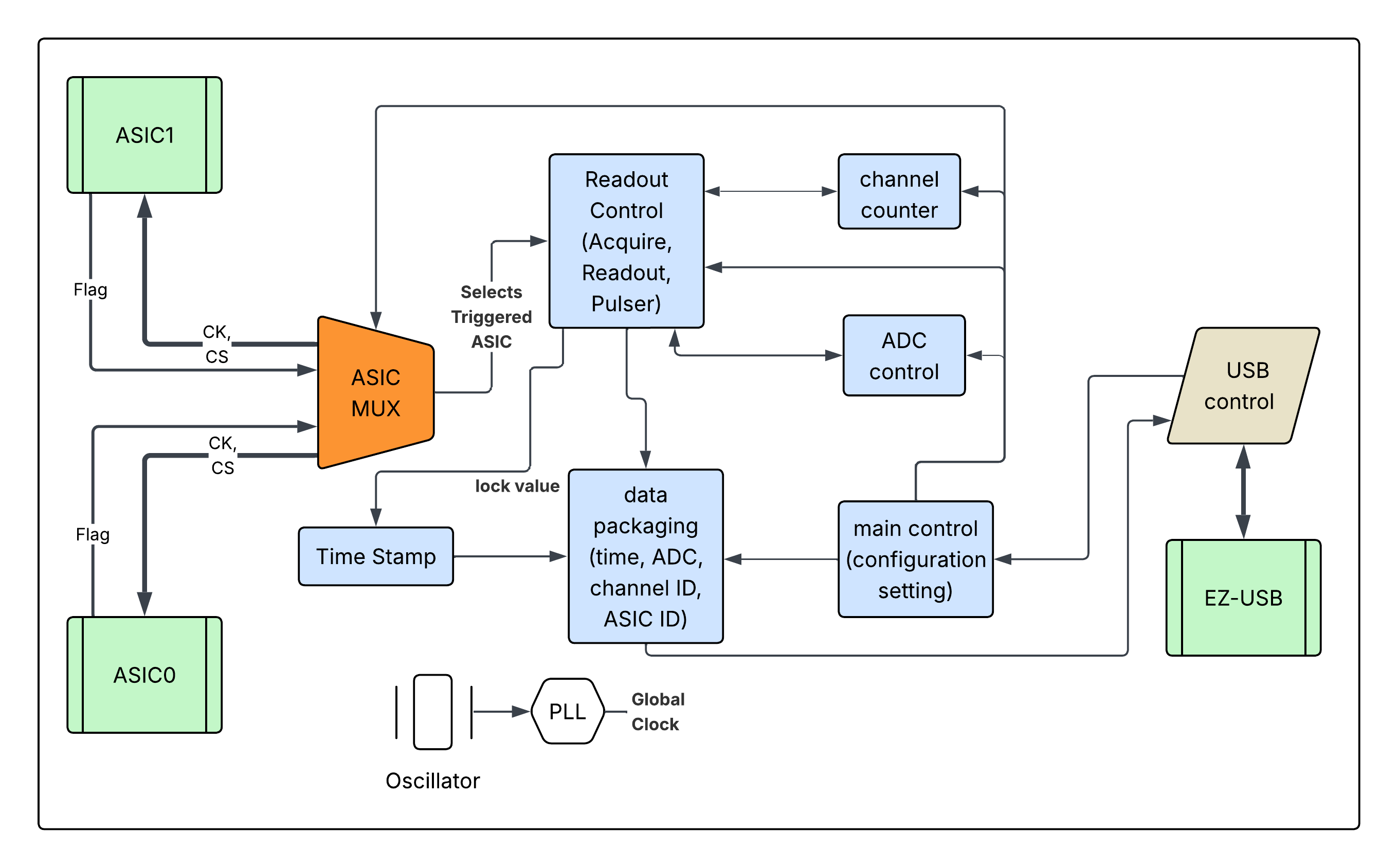}
    \caption{Top-level block diagram of the FPGA readout firmware. Blue blocks are Verilog modules inside the FPGA; ASIC0, ASIC1, the EZ-USB controller, oscillator, and PLL are external hardware components.}
    \label{fig:Readout_diagram}
  \end{subfigure}

  \vspace{1em}

  \begin{subfigure}{\linewidth}
    \centering
    \includegraphics[width=1.0\linewidth]{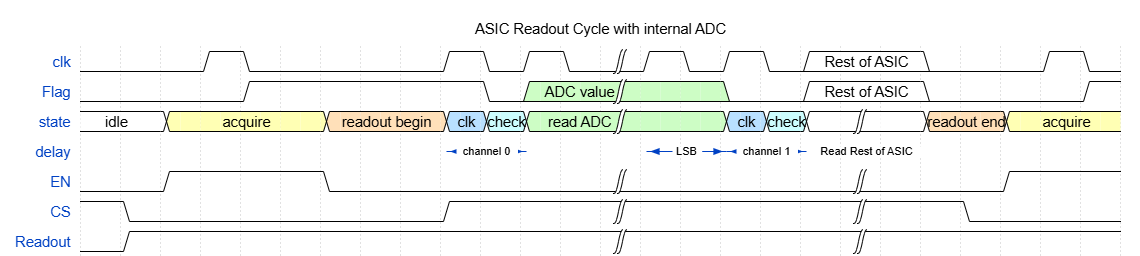}
    \caption{Schematic timing diagram of NSX1 readout through internal ADC digitization.}
    \label{fig:Int_Readout_diagram}
  \end{subfigure}
  \caption{FPGA readout firmware: (a) top-level block diagram and (b) internal-ADC readout timing.}
  \label{fig:readout_overview}
\end{figure}

The top module acts as a wrapper for the Lattice FPGA implementation. 
It instantiates the main system integration module, which instantiates all FPGA modules.
In addition, it also handles clock generation, reset management, I/O multiplexing, and board-specific I/O pins mapping. 
A system clock of 50~MHz is generated from a 150~MHz internal oscillator and on-chip PLLs. The firmware uses a power-on reset that is held for 1~$\mu$s before deassertion. 
Figure \ref{fig:Readout_diagram} shows a simplified block diagram for readout design. 

The readout module controls both acquisition and readout of the ASIC and also controls the integrated pulse generator for charge injection. In acquisition mode, the ASIC waits for an event hit with chip select (CS) low and enable (EN) high. Once an ASIC receives a hit and raises the event flag, the FPGA selects the triggered ASIC and the state machine transitions to readout mode after a programmable delay. In readout mode, the FPGA reads each channel by clocking the ASIC, receiving data over the flag pin, and sending the data to the data-control module. This process is shown in Figure~\ref{fig:Int_Readout_diagram}. An internal counter is used for the channels. Deactivated channels are not clocked, and the ADC does not read channels that do not indicate data on the flag pin during the clock high phase. The readout switch module multiplexes between the two ASICs for readout and selects based on which ASIC triggers. If both ASICs are triggered, it reads ASIC 0 first and then ASIC 1. This module also synchronizes external signals to the system clock.

There are two ADC modules: one for reading an external ADC and the other for reading the internal ADC. Both modules contain state machines for reading out the ADC. These are activated by the readout module and send their data to the data-control module. The ADC is selected by the opcode read module, which reads the bitstream that is used for ASIC programming. The internal ADC uses the flag pin for data and the same clock as the readout module. The external ADC has its own dedicated clock, convert, and data pins. The data-control module receives data from the ADC modules, a timestamp, and the channel count. It is triggered by the readout module. It packs and then sends data to the USB interface for transmission to the computer.

The opcode read module reads the opcodes from the USB interface. It receives the bitstream for programming the ASIC. This bitstream is sent to the ASIC programming module. The bitstream is also used to determine whether any channels are disabled and which ADC is being used. It also receives the delay times and the start and stop commands for readout, which are distributed to the other modules. The ASIC programming module is controlled by the opcode read module. It sends a 1024-bit SPI bitstream to the ASIC. The bitstream is 1024 bits, which is larger than the 704 bits required for programming; this allows for changes to the bitstream size for ASIC revisions without module changes, because the extra bits are shifted out. To program the ASIC, chip select and RW are held high and enable is held low. The ASIC clock is used to clock the ASIC. Data is sent over the data pin and is valid on both clock edges.

\subsection{Event Packet Format}
\label{subsec:Event_Packet_Format}

\begin{figure}[!htbp]
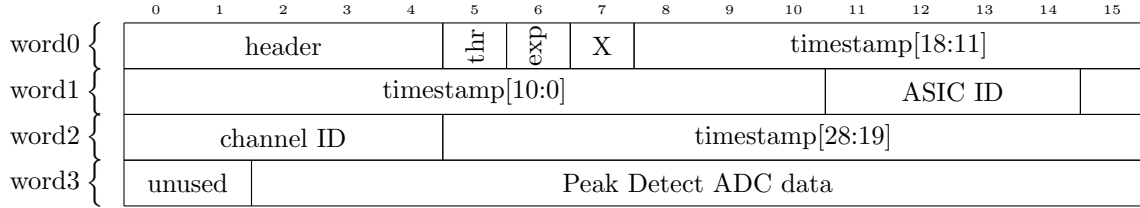

  \centering
  \begin{bytefield}[bitwidth={24pt}]{16}
    \bitheader{0-15} \\
    \begin{leftwordgroup}{word0}
      \bitbox{5}{header} & \bitbox{1}{\rotatebox{90}{thr}} &
      \bitbox{1}{\rotatebox{90}{exp}} &
      \bitbox{1}{{X}} &
      \bitbox{8}{timestamp[18:11]}
    \end{leftwordgroup}
    \\
    \begin{leftwordgroup}{word1}
      \bitbox{11}{timestamp[10:0]}&
      \bitbox{4}{ASIC ID}&
      \bitbox{1}{}
    \end{leftwordgroup}
    \\
    \begin{leftwordgroup}{word2}
      \bitbox{5}{channel ID}&
      \bitbox{11}{timestamp[28:19]}
    \end{leftwordgroup}
    \\
    \begin{leftwordgroup}{word3}
      \bitbox{2}{unused}&
      \bitbox{14}{Peak Detect ADC data}
    \end{leftwordgroup}
  \end{bytefield}
\caption{Each event is sent as four 16-bit words containing a separator, two status flags (threshold crossing, expired data), a 29-bit timestamp, the ASIC ID, the channel ID, and the 14-bit peak-detector ADC code.}
\label{fig:event_packet_format}
\end{figure}

The data-control module formats each event data into four 16-bit words (Figure~\ref{fig:event_packet_format}). Each event is made up of a 29-bit timestamp, ASIC ID, channel ID, 14-bit peak-detector ADC code, and two status flags( channel threshold crossing and expired event). The event words also have reserved patterns for marking the start of a run, the end of a run, and an inter-event separator. 
The Python-GSE parser module separates hit events from these markers by checking word1. The word1 holds values for the start-of-run, end-of-run, and separator. The parser discards the event with word1 equals \texttt{0xFFFF}.

\section{FPGA Implementation Results}
\label{sec:fpga_results}

This section reports the FPGA implementation results from synthesis, simulation, and bench measurement, following the FPGA/ASIC implementation-characterization approach in~\cite{darvishi2025pipelinestageresolvedtiming}. The first subsection covers resource utilization, static and dynamic power, and post-PAR timing results. The functional verification is performed using QuestaSim at gate level simulation. These measurements are taken at the reference operating point (50~MHz system clock, 1~MHz ASIC interface clock) unless otherwise stated. The event readout and dead-time study is performed by probing the FPGA interface pin and oscilloscope.

\subsection{Resource, Power, and Timing Results}
\label{subsec:synthesis}
\begin{table}[htbp]
    \centering

    \begin{subtable}{\linewidth}
        \centering
        \caption{FPGA Resource Utilization}
        \begin{tabular}{c|c|c|c}
        \hline
            Slices & Total LUT4s &  PFU Registers & I/O buffers\\
            \hline
            3552(22\%) & 3754(12\%) & 2049(6\%) & 59 \\
            \hline
        \end{tabular}
        \label{tab:Resources}
    \end{subtable}

    \vspace{1em}

    \begin{subtable}{\linewidth}
        \centering
        \caption{Calculated Power Consumption}
        \begin{tabular}{c|c|c}
            Static Power & Dynamic Power & Total \\
            \hline
             0.071~W & 0.02~W & 0.091~W
        \end{tabular}
        \label{tab:power_calc}
    \end{subtable}

    \vspace{1em}

    \begin{subtable}{\linewidth}
        \centering
        \caption{FPGA Timing Information}
        \begin{tabular}{c|c|c|c}
            Worst Clock Slack & Maximum Clock Frequency & Worst Critical Path & Clock to Out Max \\
            \hline
            8.704~ns  & 93.093~MHz & 10.347~ns & 13.105~ns
        \end{tabular}
        \label{tab:fpga_timing_table}
    \end{subtable}

    \caption{FPGA implementation results.}
    \label{tab:fpga_results}
\end{table}

The FPGA resource utilization in Table~\ref{tab:Resources} shows used and available resources. Slices are the logic blocks which each contain look-up tables (LUTs) and registers. The firmware uses a small fraction of the available resources, leaving headroom to scale to six ASICs per half SDD.
The immediate benefit of low resource utilization is that it allows a firmware revision where each ASIC is read out independently instead of sequentially once a hit is detected.

The Power Calculations presented in Table~\ref{tab:power_calc} are for typical operating conditions at 50~$^{\circ}$C. It assumes a 50~MHz output from the PLL and a 150~MHz output from the oscillator. The calculation was done using the Lattice Radiant Power Calculator using the Post Place and Route design. It factors in the clocks, IP blocks, and logic cells.
%
%

The worst-case FPGA timings shown in Table~\ref{tab:fpga_timing_table} relate to the signals within the FPGA. The clock-to-out timing is the difference between the clock edge and when the output actually toggles. The critical paths are the longest logic slacks. The worst clock slack is the longest delay the clock experiences; this is different from the logic delay because the clock has different routing resources. All of this contributes to the maximum clock frequency of 93 MHz. Despite it being possible to run the frequency higher, we did not do that because in our tests we could not run the ASIC clock high enough to require a faster system clock speed. The clock frequency was kept at 50 MHz.

\subsection{Functional Verification}
\label{subsec:verification}

\begin{figure}[!htbp]
  \centering
  \begin{subfigure}{\linewidth}
    \centering
    \includegraphics[width=1.0\linewidth]{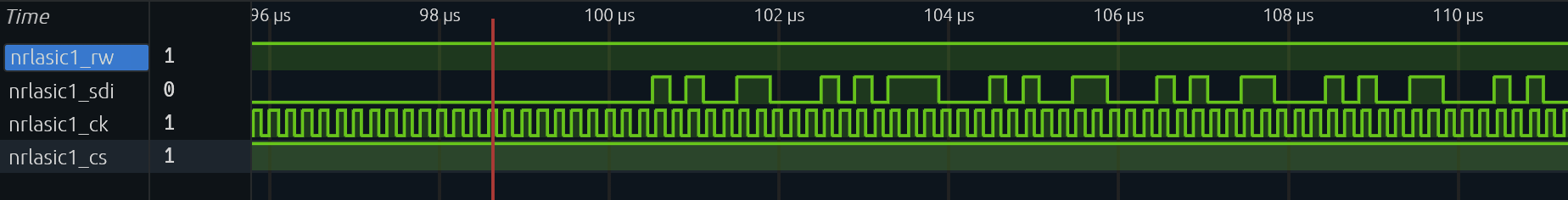}
    \caption{Gate Level Simulation of ASIC configuration over SPI}
    \label{fig:SPI_write}
  \end{subfigure}

  \vspace{1em}

  \begin{subfigure}{\linewidth}
    \centering
    \includegraphics[width=1.0\linewidth]{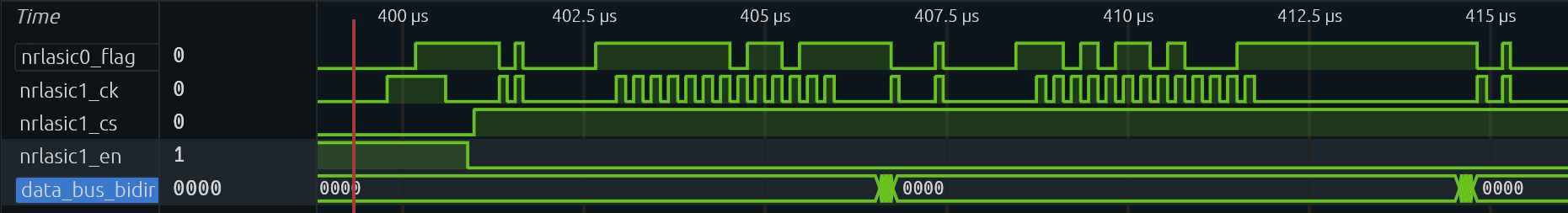}
    \caption{Gate Level Simulation of ASIC Readout using internal ADC.}
    \label{fig:Readout_simulation}
  \end{subfigure}
  \caption{Gate-level simulations of ASIC operation: (a) configuration over SPI and (b) readout using the internal ADC.}
  \label{fig:gate_level_sim}
\end{figure}

A testbench was created to verify the Verilog design for the FPGA using a gate-level simulation. For instance, a snapshot of a portion of the 1024-bit bitstream being written to the ASIC is shown in Figure~\ref{fig:SPI_write}. This is a simulation waveform of the SPI configuration write to NSX1 ASIC~1's internal registers. The FPGA is shifting the configuration bitstream into NSX1 ASIC~1, one bit per CK edge, over the SDI line. The visible window (96~$\mu$s to 110~$\mu$s, $\approx 14~\mu$s) shows the ASIC clock runs at 1~MHz, and the SDI transitions show the data pattern being shifted in.

Figure~\ref{fig:Readout_simulation} shows a gate-level simulation of the readout cycle using the internal ADC on ASIC 1. The FPGA validates channel data and latches the peak-detector serial bit from the on-chip ADC. During acquisition mode, the ASIC pulser DAC is triggered the moment the ASIC clock is pulsed if the pulser register is enabled. The testbench generates a flag in response after a delay simulating the ASIC. The flag value is set to be random in the testbench.
The individual clocks are the initial channel token clocks which validate the data in the particular channel. If the channel has no data, indicated by a low flag signal, the FPGA skips the channel's ADC readout. However, if the channel has data, the FPGA reads the ADC value by sending groups of 13 clocks to read the serial ADC output. The ADC read is started after the flag is reasserted high after a valid channel, which indicates the ADC is done with conversion.
After each ADC value is read, the data packets are sent to the host computer, this can be seen as the data bus changing in the simulation waveform. After this, it proceeds to the next channel.

\subsection{Readout Dead Time and Event Rate}
\label{subsec:deadtime}
\begin{figure}[!htbp]
  \centering
  \includegraphics[width=0.7\linewidth]{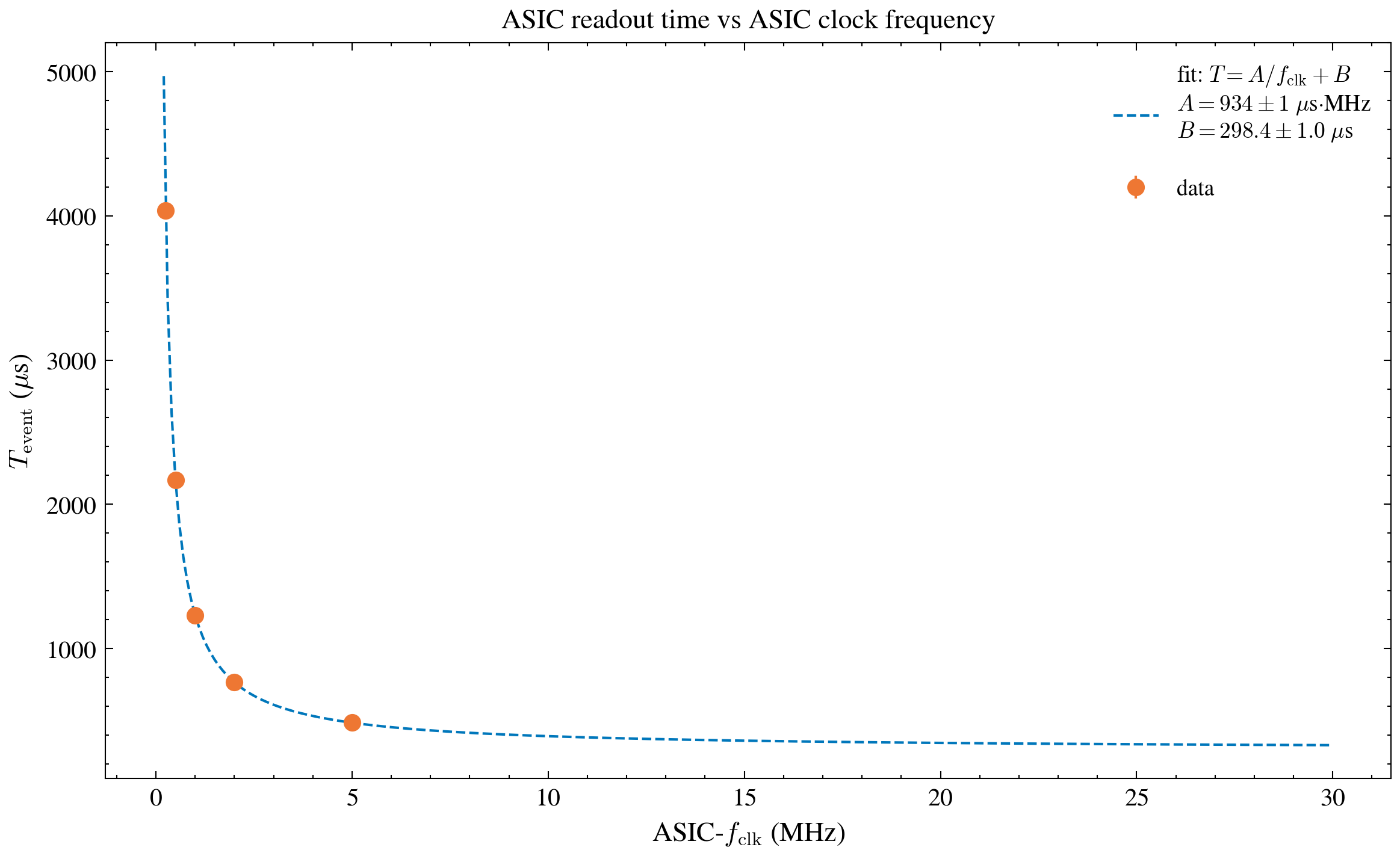}
  
    \caption{ASIC readout dead-time vs.interface clock
      $f_{\mathrm{clk}} = (50\,\mathrm{MHz})/N$.
      Markers: measured mean $\pm$ std of $T_{\mathrm{event}}$ over the
      full event readout time.
      Dashed curve: fit to $T = A/f_{\mathrm{clk}} + B$, with $A$ the
      clock-scaled cycle count and $B$ the clock-independent floor.}
    \label{fig:asic_readout_deadtime}
\end{figure}

The dead-time measurements were performed with the pulser injection frequency of $10\,\mathrm{Hz}$,
and with all 64 channels in each ASIC powered on. Figure ~\ref{fig:asic_readout_deadtime} shows the dead-time measurements.
The ASICs were configured to read out all 128 channels regardless of hit data or not.
A least-squares fit for dead time $T_{\mathrm{event}} = A/if_{\mathrm{clk}} + B$ gives $A = 934$ cycles and $B = 298\,\mu\mathrm{s}$, with $R^2 = 0.99$. 
About 934 ASIC-clock cycles are used shifting channel data per event. 
The $B = 298\,\mu\mathrm{s}$ intercept is the clock-independent floor which comprises the following delays. 
The combined clock-window and trigger-wait latency from FLAG rising edge to end of the acquire mode was $8.16\,\mu\mathrm{s}$.
The minimum spacing between chip-select (CS) and enable (EN) edges was $4\,\mu\mathrm{s}$.
The readout delay between asserting CS high and beginning ASIC clock toggling was $2\,\mu\mathrm{s}$.
A post delay of $4\,\mu\mathrm{s}$ was inserted between readout operations of two successive ASICs.
A busy delay of $200\,\mu\mathrm{s}$ was added between completion of an event and restart of acquire mode.
This dead time is the maximum possible we can have, as in most real acquisition we will have a channel multiplicity of 8-10.

The last column of Table~\ref{tab:asic_readout} lists the sustained event rate
$1/\langle T_{\mathrm{event}}\rangle$ which shows that at $if_{\mathrm{clk}} = 5\,\mathrm{MHz}$ the readout sustains ${\sim}2\,\mathrm{kHz}$.

\begin{table}[htbp]
  \centering
  \caption{ASIC readout dead time versus asic clock
  $if_{\mathrm{clk}} = 50/N$\,MHz. $\langle T_{\mathrm{event}}\rangle$ is the
  measured mean dead time from flag detection through readout completion.}
  \label{tab:asic_readout}
  \begin{tabular}{ccccc}
    \toprule
    $N$ & $if_{\mathrm{clk}}$ (MHz) &
    $\langle T_{\mathrm{event}}\rangle$ ($\mu$s) &
    $\sigma$ (ns) & Event rate (kHz) \\
    \midrule
    200 & 0.25 & 4035.0 & 365.0 & 0.25 \\
    100 & 0.50 & 2168.0 & 340.0 & 0.46 \\
     50 & 1.00 & 1231.0 & 234.0 & 0.81 \\
     25 & 2.00 &  764.5 & 142.0 & 1.31 \\
     10 & 5.00 &  486.7 &  82.0 & 2.05 \\
    \bottomrule
  \end{tabular}
\end{table}


\section{NSX1 ASIC Architecture}
\label{sec:nsx1_architecture}

\begin{figure}[!htbp]
  \centering
  \includegraphics[width=0.8\linewidth]{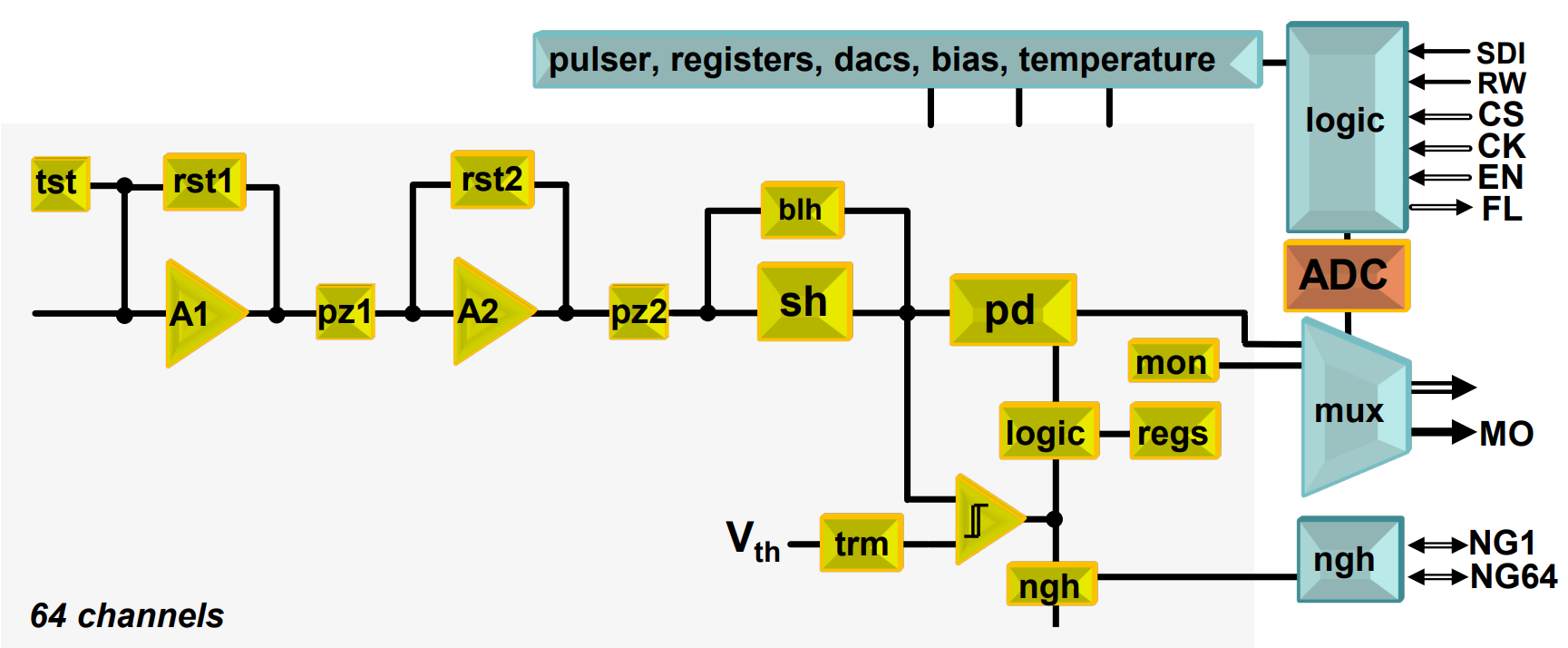}
  \caption{The NSX1 architecture showing charge-sensitive amplifier, shaper, discriminator, trim DAC, peak detector, monitor path, integrated ADC path, global DACs, and FPGA control/readout pins\cite{nsx1_datasheet}.}
  \label{fig:nsx1_architecture}
\end{figure}

The NSX1 is a low-power, low-noise front-end application-specific integrated circuit (ASIC) with 64 channels.
It is designed to detect the electronic charge signals collected by the anodes of the linear silicon drift detectors (SDDs) of the STROBE-X High Energy Modular Array (HEMA) and Wide Field Monitor (WFM) instruments~\cite{degeronimo2024frontendasicstrobexhema, nsx1_datasheet}.
Each channel has a two-stage charge-sensitive amplifier, a shaper with baseline stabilizer, a discriminator, a peak detector, and per-channel configuration logic. 
The per-channel configuration registers provide the programmable features such as charge-amplifier power-down, channel power-down, a 10~fF test capacitor, discrimination mask disables, and a 4-bit threshold trimmer. 
In addition, the ASIC provides global configuration and monitoring functionalities at the ASIC level which is shared among all 64 channels. 
It also has a programmable pulse generator that injects programmable charge through the per-channel test capacitors.
The configuration registers store the programmable settings for the ASIC's different operating modes and per-channel behavior.
Two global 10-bit DACs are used to set the threshold and the pulse-generator amplitude.
A temperature sensor reports the on-chip temperature.
The control logic block runs the digital interface that configures and reads out the ASIC.
An integrated 12-bit ADC digitizes the captured peak amplitude on chip.
An analog monitor multiplexer routes a selected channel to the analog monitor output (MO) on the SPHINX board.
The neighbor logic handles sub-threshold and inter-chip or inter-channel event readout.
The total nominal power consumption is approximately 670~$\mu$W per channel.
A simplified block diagram of the NSX1 is shown in Figure~\ref{fig:nsx1_architecture}.

The NSX1 provides the additional global configurations such as a high-resolution mode, detect at first peak, trigger all channels at threshold crossing, and sample all channels at first peak found. 
Using the high-resolution mode increases the power consumption by approximately 200~$\mu$W per channel.
A programmable current source injects approximately 1~pA or 20~pA at the input to emulate the sensor leakage current.

The front-end signal processing for each channel is shown inside the gray box in Figure~\ref{fig:nsx1_architecture}. 
Each channel begins with a two-stage charge-sensitive amplifier (CSA) followed by a shaper, discriminator, and peak detector circuit. 
The CSA provides two selectable front-end gain configurations of 375~mV/fC and 750~mV/fC.
It also provides a continuous, leakage-adaptive mirror reset with a pole-zero cancellation circuit. 
The CSA output is shaped using a third-order shaper with peaking times programmable to 250~ns, 500~ns, 1~$\mu$s, or 2~$\mu$s. 
It also includes baseline stabilization, which is referenced to an on-chip bandgap reference (BGR), and sets the shaper output baseline at approximately 150~mV.
The ASIC includes a discriminator (with a fixed hysteresis of a few millivolts) and its threshold is set by a global 10-bit DAC and a per-channel 4-bit trim DAC. 
When the discriminator detects a threshold crossing, the peak detector circuit captures the maximum amplitude of the shaped pulse, and stores the peak value in an analog memory buffer.

The ASIC readout operates in two distinct phases: acquisition and readout.
During acquisition all powered-on channels are sensitive to a hit, and when a peak is detected, the ASIC asserts the event flag (FL) high after a delay of approximately 400~ns to indicate that one or more valid events are present.
The FPGA then stops acquisition and enters readout mode.
During the readout phase, an internally generated token advances through the channels on successive rising edges of the clock (CK), and the number of clock cycles represents the channel ID.
The FL line carries the event availability, threshold crossing, and channel data. The captured peak amplitude is digitized internally by the integrated ADC, and the 12-bit code is then shifted out serially on the FL line. For details, refer to Section~\ref{sec:hdl_methodology}.

\section{NSX1 ASIC Characterization}
\label{sec:asic_characterization}

For the characterization methodology, the NSX1 is treated as the device under test (DUT) and the FPGA and Python-GSE software as the calibrated measurement instrument.
For the measurement procedure, the on-chip pulse generator is used to drive the per-channel test capacitor for charge injection, and the resulting peak amplitudes are digitized either by the external AD7980 ADC or by the integrated 12-bit ADC.
The ASIC integrates an analog monitor to route internal analog signals to the dedicated buffered analog output (MO). 
For instance, ASIC-level reference voltages and DAC voltage, and channel-level analog outputs and trimmed threshold.
All measurements using the analog monitor are performed using a PicoScope 5244D MSO with 200~MHz bandwidth with time resolution of 16~ns.
All configurations for a particular run are always saved in a JSON file alongside HDF5 files which contain event packets and metadata.

\subsection{Digital Interface, Operating Modes, and Timing Verification}
\label{subsec:digital_interface_timing}

\begin{figure}[!htbp]
  \centering
  \includegraphics[width=0.9\linewidth]{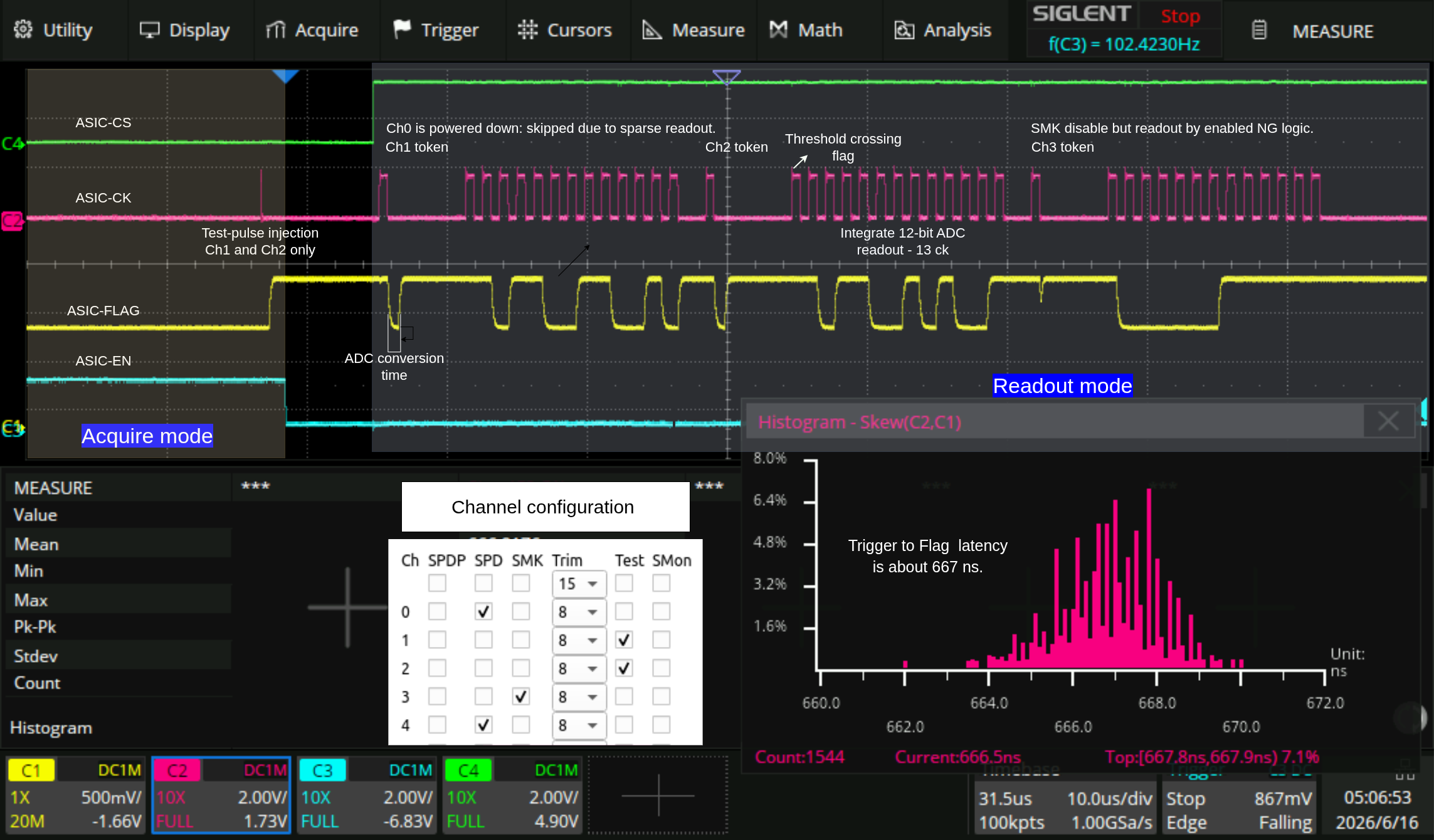}
  \caption{Integrated ADC readout timing: acquire mode (flag set on a hit) and sparse readout (13 clocks/channel, token per channel; powered-down channels skipped). Inset: trigger-to-flag latency (~667 ns).}
  \label{fig:digital_interface_timing}
\end{figure}

The digital lines (EN, CS, CK, FL, RW, and SDI) for both ASICs are probed with an oscilloscope during configuration, acquisition, and readout phases. The timing measurements are taken with the FPGA system clock of 50~MHz and ASIC-interface clock of 1~MHz.
Figure~\ref{fig:digital_interface_timing} shows an oscilloscope capture of the interface signals during the acquire and readout phases.
A test pulse is used to inject charge into channels 1 and 2 only (channel 0 is powered down). 
During acquire mode (left), the ASIC waits for a hit, and when an event is detected, the event flag (FL) is asserted. 
The ASIC then enters readout mode (right) after a programmable delay.
During readout mode, all the powered-up and valid channels are read out where each channel's 12 digital bits are shifted out serially on FL in order of MSB first to LSB last.
The first clock data represents whether the channel crossed the threshold or not, useful when the neighbor function is enabled.
The channel configuration (power, mask, and test-pulse settings) used for this measurement is shown in the inset white block.
The inset histogram gives the measured trigger-to-flag latency, which is about $667 \pm 1$~ns.

\subsection{Pulser DAC Linearity}
\label{subsec:pulser_charge_response}

\begin{figure}[!htbp]
  \centering
  \begin{subfigure}[t]{0.49\linewidth}
    \centering
    \includegraphics[width=\linewidth]{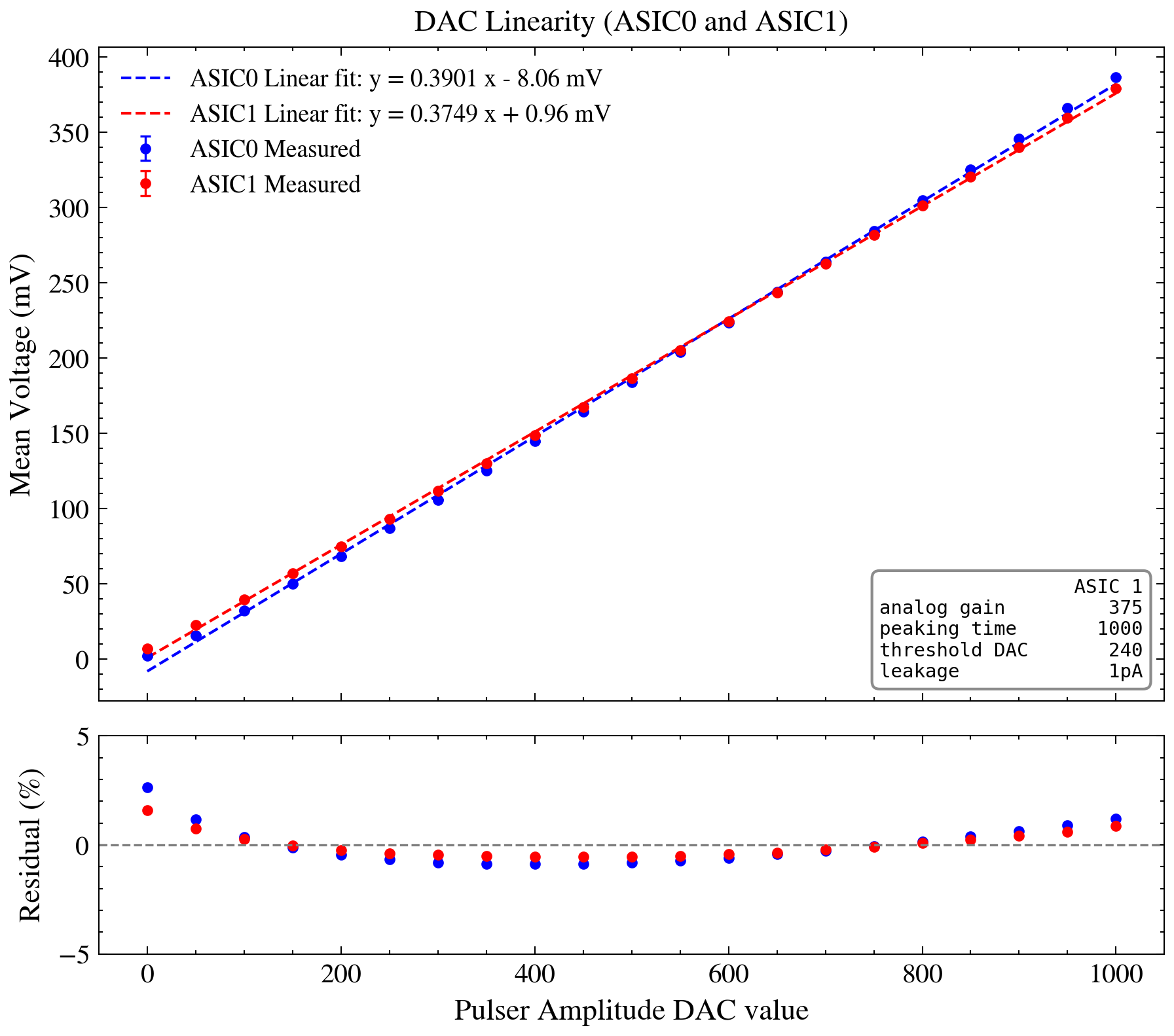}
    \caption{Pulser DAC output as a function of input Pulser DAC codes(voltage domain). The residual is expressed as a percentage of the full-scale voltage (INL).}
    \label{fig:pulser_charge_response}
  \end{subfigure}
  \hfill
  \begin{subfigure}[t]{0.49\linewidth}
    \centering
    \includegraphics[width=\linewidth]{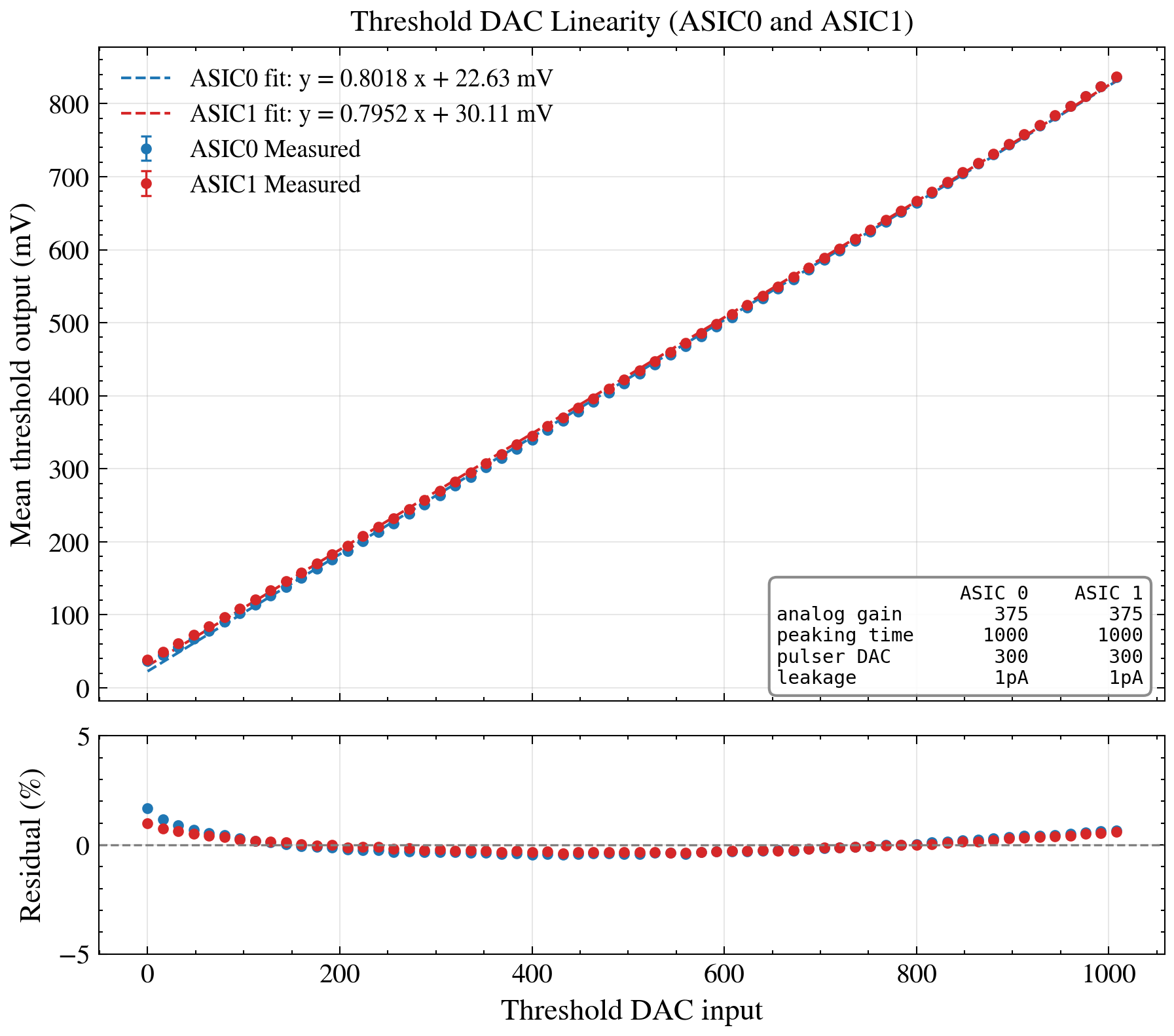}
    \caption{Threshold DAC output as a function of input threshold DAC codes(voltage domain).
    The residual is expressed as a percentage of the full-scale voltage (INL).}
    \label{fig:threshold_trim}
  \end{subfigure}
  \caption{Internal pulser and threshold DAC linearity.}
\end{figure}

The NSX1 contains a 10-bit pulser DAC to set a voltage level of an internal test pulse that drives the test capacitor for each channel. 
The response of the pulser DAC as a function of input DAC code is measured as an analog voltage at the analog monitoring output of the NSX1 via a PicoScope.
Figure~\ref{fig:pulser_charge_response} shows the 10-bit DAC calibration for the internal test pulser. 
The data is fitted with a linear calibration function $y = mx + b$ for DAC values greater than 5. 
A non-linearity in the lower DAC code region is due to the limit of the external analog buffer when operating close to 0V.  

\FloatBarrier


\subsection{Threshold DAC and Trim Verification}
\label{subsec:threshold_trim}
To set the discriminator voltage level, NSX1 contains one 10-bit threshold DAC shared to all channels and 4-bit trim DAC
for each individual channel. 
The trim DAC is to fine-tune the threshold level so that the effective threshold level for accross channels is uniform.
The trimmers operate only in one direction by lowering the global threshold spread per channel.

Figure~\ref{fig:threshold_trim} shows the 10-bit DAC calibration for the discrimination threshold level. 
The data is fitted with a linear calibration function $y = mx + b$ for DAC values greater than 5. 
In the lower DAC code region, there is nonlinearity, but this is due to the inefficiency of the external analog buffer at the MO. 
Nonetheless, the DAC response is linear in the region above the baseline up to the full DAC range.

\begin{figure}[!htbp]
  \centering
  \includegraphics[width=0.7\linewidth]{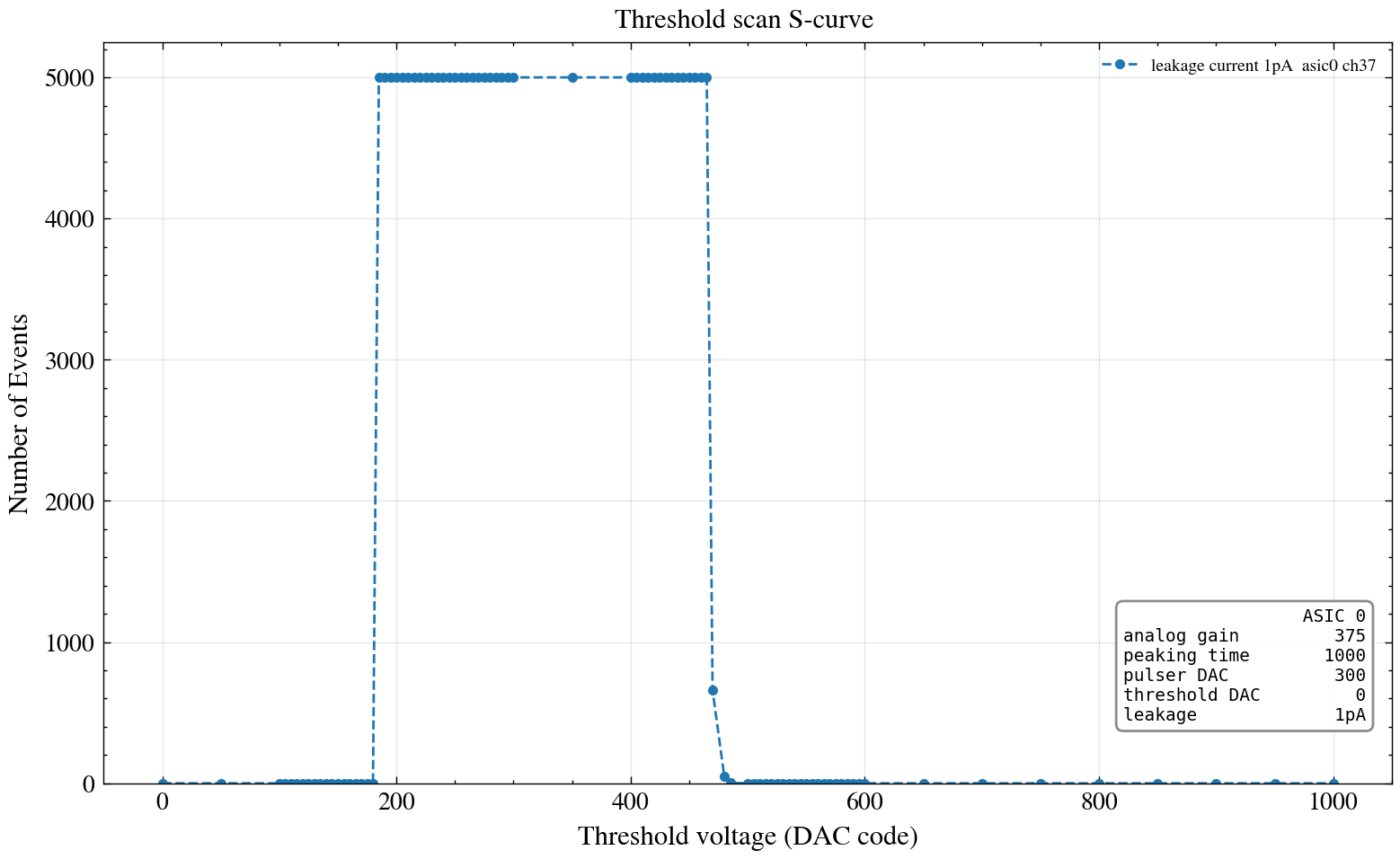}
\caption{Threshold scan for ASIC~0, ch~37: hit fraction versus threshold DAC code for a fixed injected test pulse (DAC 300). }
  \label{fig:threshold_s_curve}
\end{figure}

The threshold scan is performed by injecting a constant test pulse into a single channel (ASIC~0, channel 37) while the discriminator threshold is stepped across its full DAC range.
At each threshold setting, hit fraction is counted and plotted again the threshold DAC input.
Figure~\ref{fig:threshold_s_curve} shows the threshold scan where the channel counts every pulse only within a band of threshold codes (about 180 to 500) and counts zero outside it suggesting threshold sits between the baseline and the pulse peak. 
The shape edge implies that the noise is quite small relative to the pulse.

\subsection{Channel response: linearity, and saturation}
\label{subsec:pulser_gain_dynamic_range}

\begin{figure}[!htbp]
  \centering
  \includegraphics[width=0.49\linewidth]{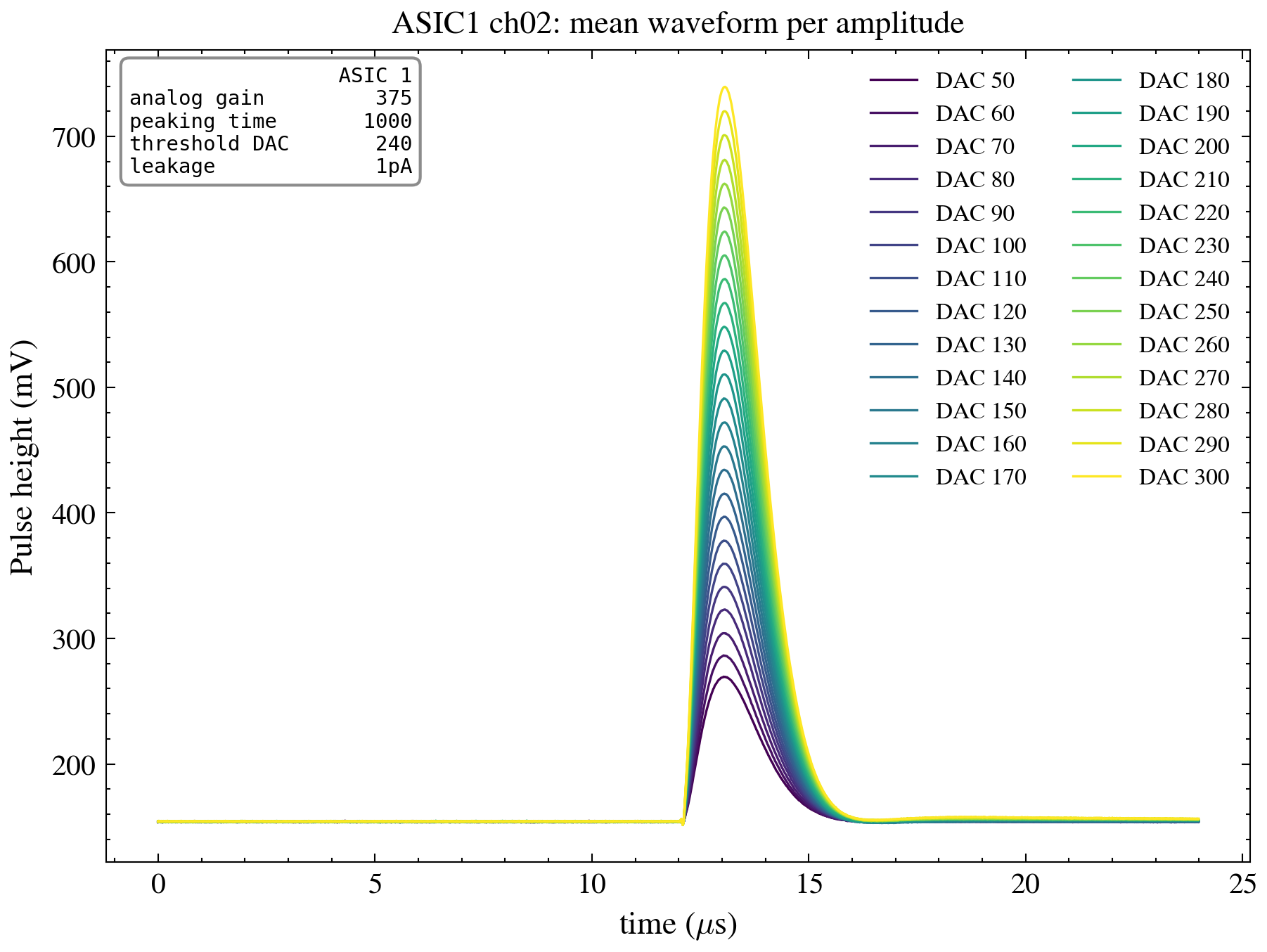}
  \hfill
  \includegraphics[width=0.49\linewidth]{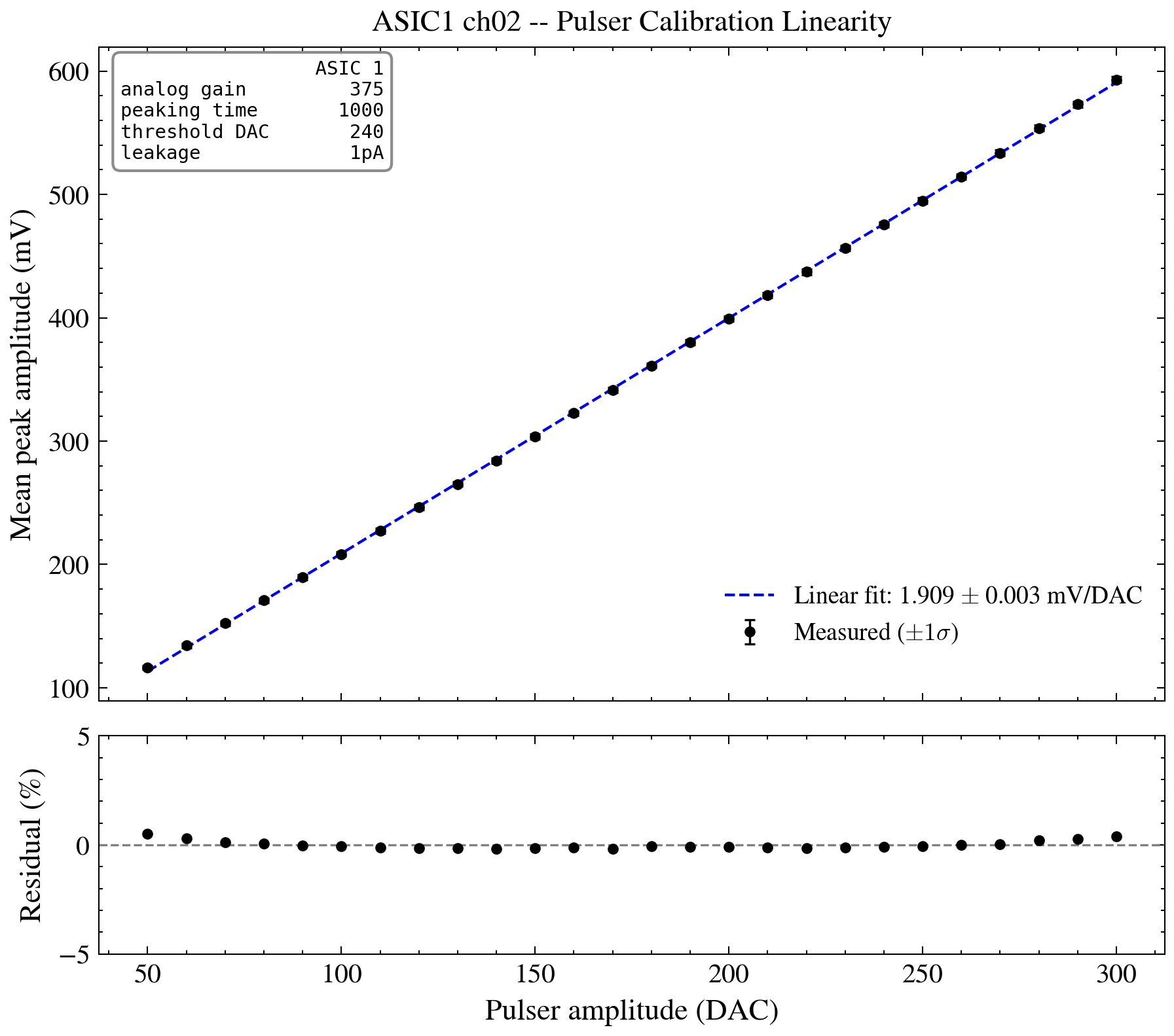}
  \caption{(Left): Mean waveform of the shaper output per amplitude for ASIC~1, channel~2. (Right): Top: Linearity of the pulser response for ASIC~1, channel~2. Bottom: The deviation as a percentage of the full-scale voltage (INL).}
  \label{fig:pulser_response_linearity}
\end{figure}

The shaper pulse response at different input test amplitudes (derived from the internal pulser calibration ~\ref{subsec:pulser_charge_response}) for ASIC~1 channel 2 is shown in Figure~\ref{fig:pulser_response_linearity} for a nominal peaking of 1 $\mu$s, threshold DAC code of 240, with 1~pA leakage emulator. 
The front-end input is loaded with a capacitor of 53~fF (calculated). 
A test-pulse sweep is performed through DAC code in the range of 50 to 300. Figure~\ref{fig:pulser_response_linearity} on the left shows the average amplitude response measured as the peak amplitude versus pulser DAC setting. 
The peak amplitude scales with the injected test charge while keeping a constant shaping response.
The right plot shows the linear response of mean raw peak height as a function of input test amplitude. 
A linear fit is performed over the region, and the residual nonlinearity is reported as the deviation from that fit. 
The deviation as a percentage of full-scale range is reported in the bottom panel and
found to be within $\pm$0.5\% with in the lower end of DAC ranges.

\begin{figure}[!htbp]
    \centering
  \includegraphics[width=0.7\linewidth]{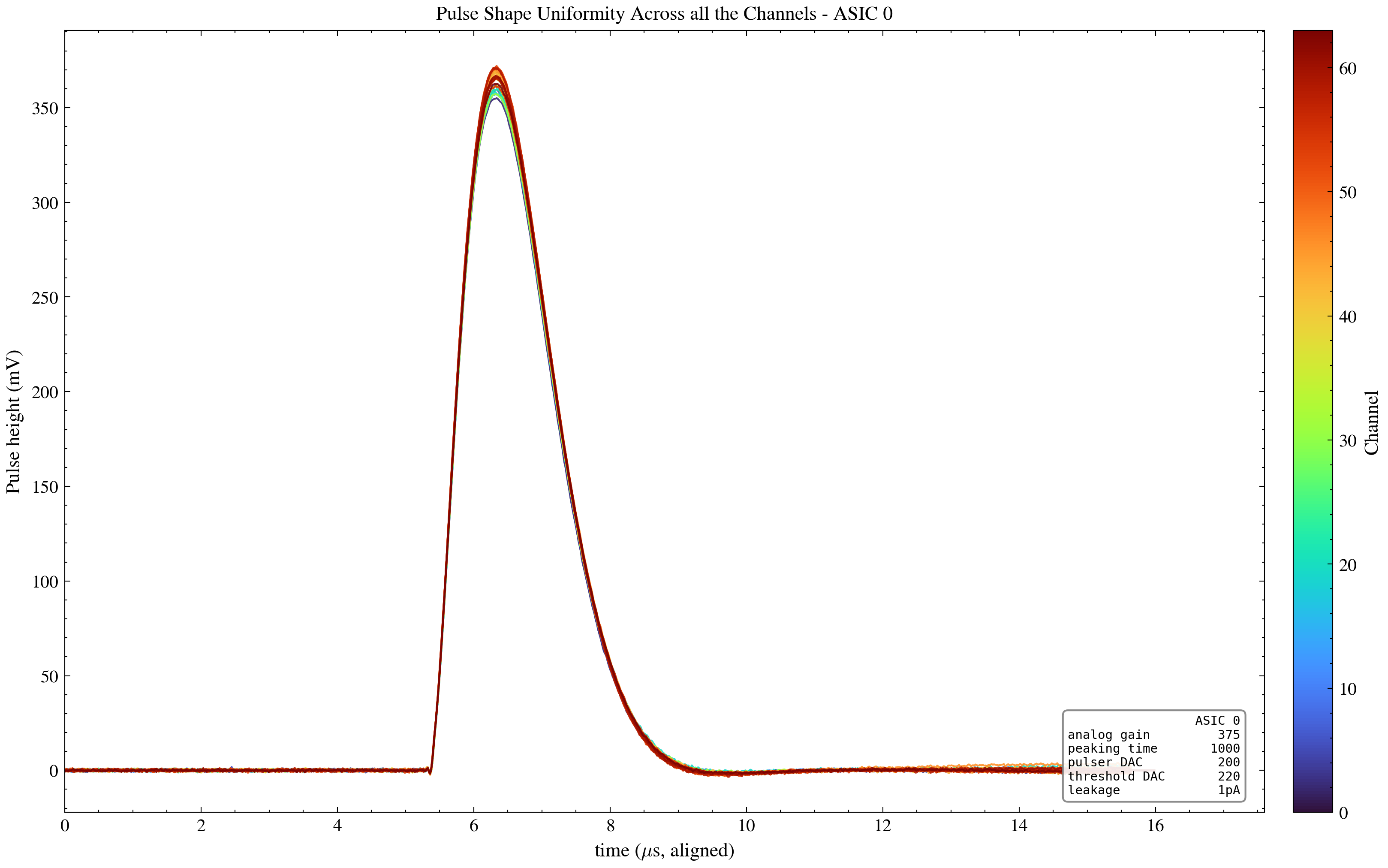}
  \includegraphics[width=1\linewidth]{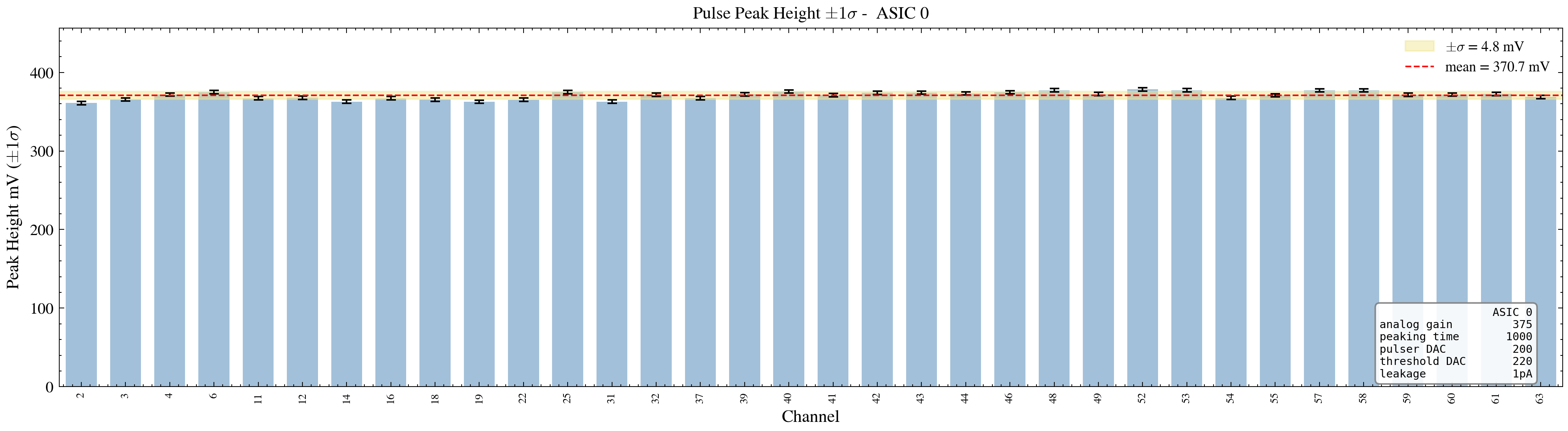}
  \caption{(Top): Mean waveform of the shaper output per channel for ASIC~1. (Bottom): Peak height uniformity of the pulser response for ASIC~1 across all the channels.}
  \label{fig:pulser_response_accross_channel}
\end{figure}

Channel-to-channel uniformity of the front-end response is checked by injecting a fixed-amplitude test pulse into every channel and recording the averaged shaper-output waveform through the analog monitor using PicoScope. 
The time-aligned mean waveform of each channel is overlaid in Figure~\ref {fig:pulser_response_accross_channel} (top). 
The peak height extracted from each waveform is summarized in Figure~\ref{fig:pulser_response_accross_channel} (bottom), with a mean of 370.7~mV and a channel-to-channel spread of 4.8~mV ($\sim$1.3\%). 
One sigma error bar on each channel is its event-to-event spread, which is well below the channel-to-channel variation. 
The uniform shape and the $\sim$1\,\%amplitude spread confirm that the analog gain and shaping are consistent across the channels at the fixed gain, peaking time, and injected amplitude.

\subsection{Analog Gain, Peak Time, and Pulse Shape}
\label{subsec:analog_gain_pulse_shape}

\begin{figure}[!htbp]
  \centering
  \includegraphics[width=0.7\linewidth]{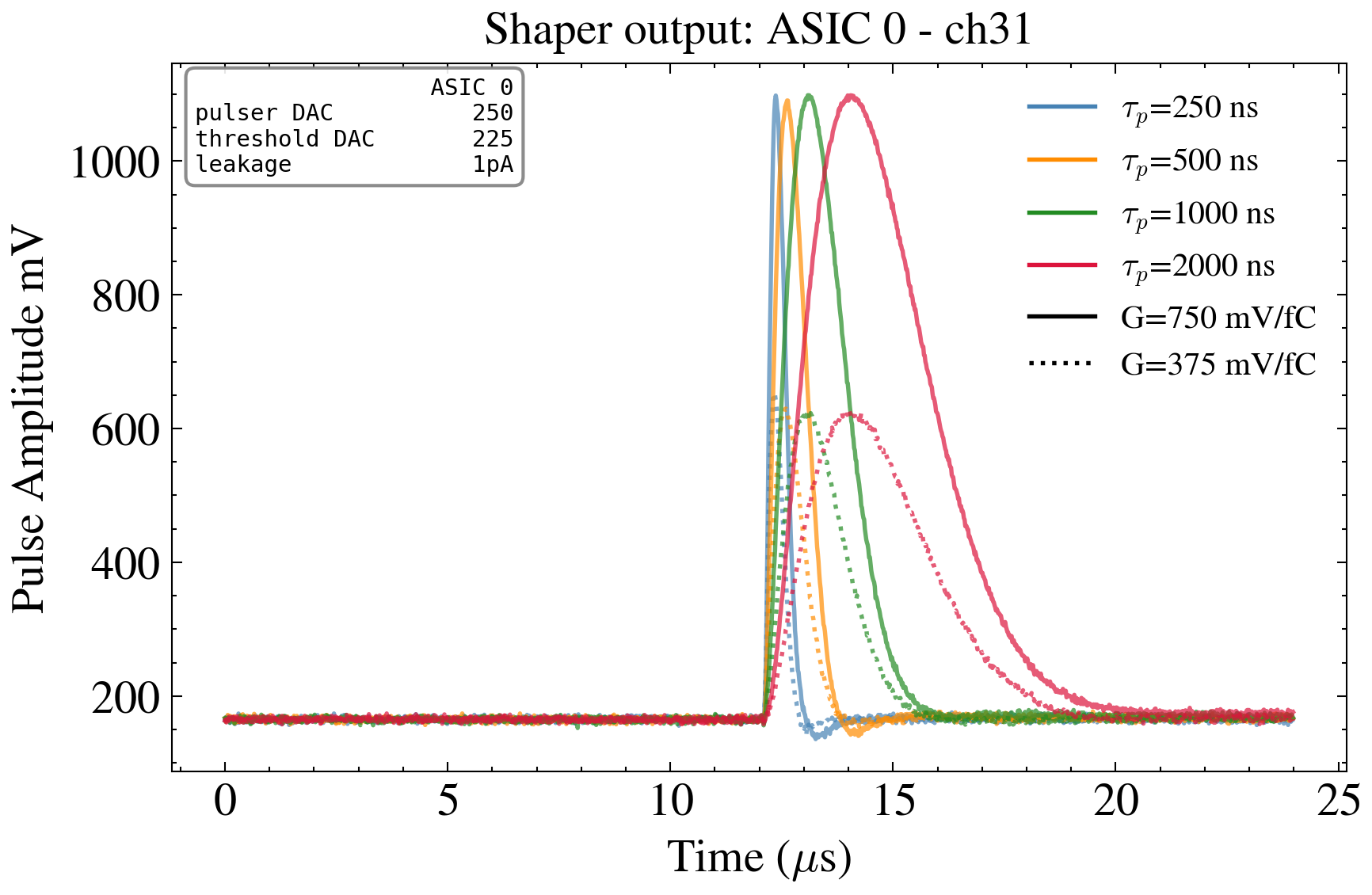}
  \caption{Analog gain, peak-time, and pulse-shape characterization. An average shaper response for a fixed test-pulse injection.}
  \label{fig:analog_gain_pulse_shape}
\end{figure}

To characterize the front-end ASIC's programmable controls such as analog gain and peaking time, we measure shaper response of each channel by injecting a test-pulse. 
Figure~\ref{fig:analog_gain_pulse_shape} shows the average shaper output of ASIC~0, channel 31 for two programmable analog gains (375 and 750~mV/fC) and four programmable peaking times(250~ns, 500~ns, 1~$\mu$s, and 2~$\mu$s).
The width of a output pulse increases from 250~ns to 2~$\mu$s, and for each peaking time, the high gain pulse height is approximately twice the pulse height of low gain. 
All pulses rise from the baseline line of $\sim$172~mV and return to the baseline with a semi-Gaussian unipolar shape.

\subsection{Sparse readout and Neighbor Logic}
\label{subsec:neighbor_sparse_common_mode}

We verified the NSX1's event-driven readout logic by controlled test-charge injection. 
Out of 64 channels, we randomly powered down some channels and randomly masked other channels.
All the masked and powered-down channels were injected with test charge along with other randomly selected channels.
The readout output verified that the powered-down and masked channels were not read out while the rest of the other channels were read out. It returns the correct channel identity. 

To verify the neighbor logic, we repeat the sparse readout configuration but this time we enabled the neighbor channel readout logic for a couple of channels, including inter-chip channels. 
The output verified that all the adjacent channels are read out correctly even though those channels shaper output are below the threshold for each and every neighbor logic enabled channel. Figure ~\ref{fig:digital_interface_timing} shows the additional information about timing diagram of sparse readout and neighbor logic verification.

\begin{table}[ht]
  \centering
  \caption{ASIC~1: sparse-readout test-pulse run with neighbor logic enabled.}
  \label{tab:asic1_sparse}
  \begin{tabular}{@{}l p{12.7cm} c@{}}
    \toprule
    Group & Channels & Count \\
    \midrule
    Injected, fired    & 2, 6, 13, 18, 23, 29, 34, 39, 45, 50, 55, 61 & 12 \\
    Neighbor-readout   & 0, 1, 3, 5, 7, 11, 15, 17, 19, 22, 24, 28, 30,33, 35, 38, 40, 44, 46, 49, 51, 53, 56, 60, 62 & 25 \\
    Baseline pick-up & 48 & 1 \\
    Live, never read   & 4, 8--10, 16, 20, 21, 25, 27, 31, 32, 36, 37, 41--43, 52, 57--59, 63 & 21 \\
    Powered down       & 12, 14, 26, 47, 54 & 5 \\
    \midrule
    Total              & & 64 \\
    \bottomrule
  \end{tabular}
\end{table}

\begin{figure}[!htbp]
  \centering
  \includegraphics[width=1\linewidth]{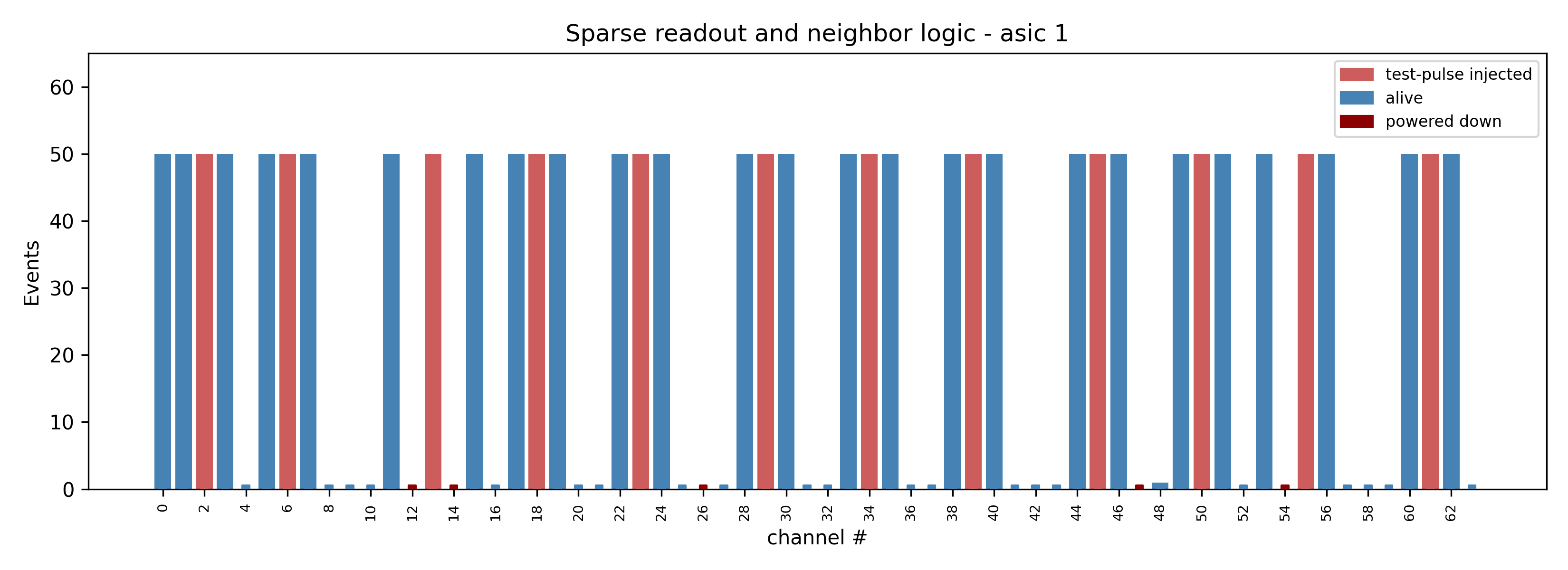}
  \caption{Sparse readout and neighbor logic bar plot}
  \label{fig:asic1_sparse_bar}
\end{figure}

Table~\ref{tab:asic1_sparse} and the occupancy bar chart in Figure~\ref{fig:asic1_sparse_bar} verify the sparse readout and neighbor logic for ASIC~1.
Out of the 64 channels, five are powered down and the rest are alive. The twelve light-red channels are injected with the test pulse.
Every alive channel adjacent to a light-red channel is read out by the neighbor logic, even though it stays below threshold.
When a channel is powered down, the neighbor logic skips it and treats the next active channel as the neighbor.
Channels that are neither injected nor a neighbor of an injected channel are not read out, which is the expected behavior of the sparse readout logic.
To verify the inter-chip neighbor logic, we deliberately inject the last enabled channel of ASIC~0, which then reads out the first channel of ASIC~1, demonstrating proper inter-chip trigger communication.


\subsection{Noise Studies}
\label{subsec:enc_studies}

\begin{figure}[!htbp]
  \centering
  \includegraphics[width=1\linewidth]{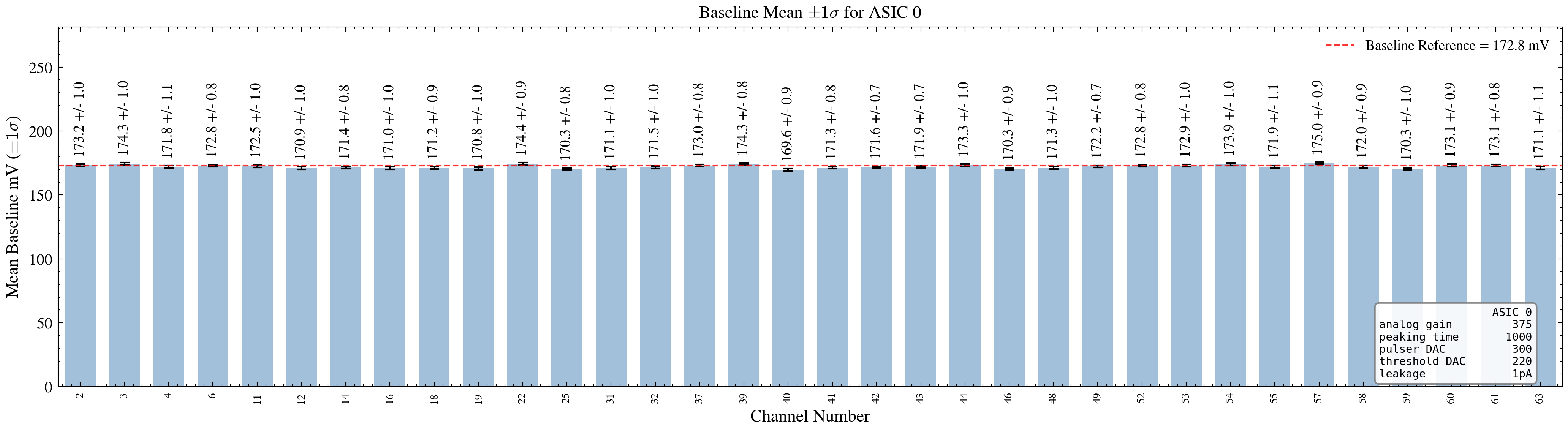}
  \caption{Per-channel baseline mean and event-to-event RMS for ASIC~0.}
  \label{fig:enc_rms_accross_channels}
\end{figure}

The front-end noise is reported as the baseline root-mean-square (RMS). The baseline RMS is measured per channel from the shaper output in the absence of injected test charge and across all channels. 
The mean baseline of each channel is summarized in Figure~\ref{fig:enc_rms_accross_channels}, with a mean of 172.8 mV and a channel-to-channel spread of 1.6 mV ($\sim$1\%). 
One-sigma error bars on each channel represent the event-to-event baseline RMS, which is below the channel-to-channel variation.

\section{Python-GSE Software and Analysis Pipeline}
\label{sec:python_gse_pipeline}

Python-GSE is a python based multiprocessing data acquisition framework.
It configures the hardware, runs the data acquisition process, and saves the data for offline analysis. 
During start-up, the GSE reads a JSON file to load the ASIC registers setting (such as gain, peaking time, threshold trim, pulser, neighbor logic, and channel masks) as well as the FPGA control registers.
The start of a DAQ run commits the configuration setting and then creates separate threads for USB-FIFO reading, packet parsing, and HDF5 writing are started.

During the data acquisition run, the USB-FIFO reading threads continuously pull 512-byte packets from the USB FIFO.
Once a buffer is received, the raw buffer is recorded in the HDF5 file along with its UTC time. 
The buffer is also sent to the Packet Parser module where it is parsed to reconstruct the channel data from the ASIC ID, channel ID, hit timestamp, peak amplitude ADC value, and flag for threshold crossing using every 4-word sequence.
This parsed data is sent both to the HDF5 writer and the Hits handler which adds it to a spectral display and rates table for the live monitoring.
The HDF5 file records the configuration settings, raw buffers, UTC time of each received buffer, and parsed event data for offline analysis.  
When the run is stopped, data acquisition sends a stop command to the FPGA. Then it closes the open threads and becomes ready for a new run.

\section{Conclusion and Outlook}
\label{sec:conclusion_outlook}

We have developed and validate an FPGA-based readout system for NSX1-ASIC characterization using Lattice Certus-NX FPGA.
The firmware drives the full ASIC interface, sequences the readout phases, and pushes the data to host PC over the USB-FIFO link.
The firmware uses a small fraction of the available resources 3552 slices (22\%), 3754 LUT4s (12\%), 2049 registers (6\%), 59 I/O buffers and provide the headroom to include 6 ASICs per half-SDD.
The firmware is verified through Gate-level QuestaSim simulations. The typical timing diagrams are presented for correct SPI configuration writes (704 out of 1024-bit bitstream) and internal-ADC readout sequencing in Section~\ref{subsec:verification}.
The dead time and sustained event is measured to be  ~2~kHz at a 5~MHz ASIC interface clock and 50~MHz FPGA system clock.
The Python-GSE software, a user interface software, is developed to configure the hardware, control the data acquisition, record the data to HDF5 file. It also provides the live data monitoring via spectral and rate display GUI windows.

The NSX1 configuration and operation modes have been verified via probing the digital interface pins. 
The device has been functionally characterized using the on-chip test-pulse generator.
The ASIC configuration and operation modes have been verified. 
The measurement of trigger-to-flag latency is found to be $\sim$667$\pm 1$~ns.
The pulser DAC responses are measured and verified to be linear using internal test-pulse injection. 
The 10-bit threshold along with a 4-bit per-channel trim DAC responses were measured
Sparse and neighbor logic readout were verified, including inter-chip and powered-down channels.

Future work targets the next version NSX1 ASIC characterization as well as scaling the number of ASICs to 12. It also involves redesigning the firmware and reading out the ASICs independently. To synchronize the detector and/or ASIC readout, an external pulse-per-second (PPS) signal implementation will be done in the next version of the firmware. 
The ASIC will be characterized with a prototype WFMino-SDD X-ray source irradiation. Radiation and single-event-upset (SEU) evaluation test will be performed.



\section*{Acknowledgments}

This work was supported by the 2023 NASA EPSCoR program and the APRA project at the Naval Research Laboratory (NRL). We also acknowledge the COSI experiment team for providing the FPGA firmware (Verilog) and the foundational electronics design framework.


\bibliography{report} 
\bibliographystyle{spiebib} 
\FloatBarrier 
\end{document}